\documentclass{article}

\usepackage[english]{babel}

\usepackage[letterpaper,top=2cm,bottom=2cm,left=3cm,right=3cm,marginparwidth=1.75cm]{geometry}

\usepackage{amsmath}
\usepackage{graphicx}
\usepackage[colorlinks=true, allcolors=blue]{hyperref}

\title{Adversarial and Safely Scaled Question Generation}
\author{Sreehari Sankar, Zhihang Dong}

\begin{document}
\maketitle

\begin{abstract}
  Question generation has recently gained a lot of research interest, especially with the advent of large language models. In and of itself, question generation can be considered "AI-hard", as there is a lack of unanimously agreed sense of what makes a question "good" or "bad".  In this paper, we tackle two fundamental problems in parallel: on one hand, we try to solve the scaling problem, where question-generation and answering applications have to be applied on a massive amount of text without ground truth labeling. The usual approach to solve this problem is to either downsample or summarize. However, there are critical risks of misinformation with these approaches. On the other hand, and related to the misinformation problem, we try to solve the 'safety' problem, as many public institutions rely on a much higher level of accuracy for the content they provide. We introduce an adversarial approach to tackle the question generation safety problem with scale. Specifically, we designed a question answering system that specifically prunes out unanswerable questions that may be generated, and further increases the quality of the answers that are generated. We build a production-ready, easily-plugged pipeline that can be used on any given body of text, that is scalable and immune from generating any hate speech, profanity or misinformation. Based on the results, we are able to generate more than six times the number of the quality questions generated by the abstractive approach, and with a perceived quality being 44\% higher, according to a survey to 168 participants.
\end{abstract}

\section{Introduction}

Question generation is the task of generating questions given context and keywords. Recently, with the advent of large language models, question generation has been applied to a wide variety of domains, the most sensitive of which is probably question generation from children's textbooks and stories for educational purposes \cite{zhao2022educational,lee2018automatic}. Apart from these, there have been a lot of instances where question generation is applicable to a large user base. When the data size is sufficiently large, human supervision of these generative systems is not feasible. In order to scale such question generation systems, abstractive summarization is one of the most widely used techniques \cite{10.1145/3442381.3449892, duan-etal-2017-question,du2017learning, nwafor2021automated,agarwal2022deep}.

\begin{figure}[h]
  \centering
  \includegraphics[width=\linewidth]{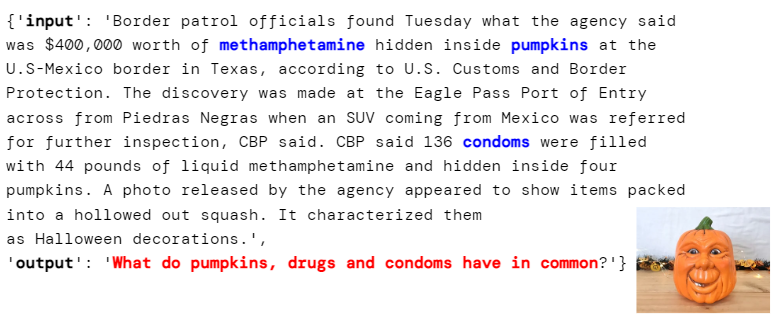}
  \caption{Example of Abstractive Model Failed to Work}
  \label{fig:res}
\end{figure}

With that being said, let us invite our readers to examine one decently interesting examples of how state-of-the-art abstractive models (\texttt{google/pegasus-xsum}) can be shockingly wrong. For Figure~\ref{fig:res}, to the best of the knowledge of the two authors, we have failed to discover the commonality between pumpkins, drugs and, especially, \textit{condoms}. This question is left for future research. However, what's more dangerous than comparing pumpkin against condom is that abstractive summarization has been shown to create misinformation \cite{shu2020combating, esmaeilzadeh2019neural}. Misinformation is dangerous to the society, particularly with the rise of fabricated misinformation \cite{wang2018era, angeline2020can}. Misinformation destroys trust between people and institutions, particularly in the area of public health \cite{swire2019public}, political engagement \cite{jerit2020political} and science \cite{west2021misinformation}.

Modern question generation frameworks are usually following a generative approach, where the input is the "context" from which the questions are asked -- if the process is "answer-guided", then additional keywords follow suit. One caveat behind this is many modern AI-generated question answering models may produce false information. As these question answering frameworks are massively scaled, we would soon reach to a point when human monitoring becomes ineffective. Even if human-in-the-loop monitoring is realistic, there were still mixed results for the human-in-the-loop or crowdsourced truth verification during the influx of information and breaking news \cite{soprano2021many, roitero2021can}. There are certain scenarios in the AI-generated question-answering applications where absolute truths must be hold: consider using such applications in the event of election, legal inquiries and health documents, we must generate answers from and only from the traceable texts.

In this paper, we tackle the problem of scaling without inviting the risk of creating misinformation. To do this, we avoid the use of summarizers to downsample the input text. Instead we devise a 'maximal generation' approach. We also propose an adversarial approach to question generation, and results show that the generated questions from an adversarial system are more than 40\% better than a vanilla abstractive question generation system. As part of our adversarial system, we integrate a question answering system in addition to the question generator to filter out 'bad' questions. We use state of the art question answering system trained on SQuAD v2.0 \cite{rajpurkar2018know}, which contains negative examples. We also show results from a model trained on SQuaD v1.1 for further understanding the effectiveness of our adversarial approach when the training resources are limited. We are pursuing question generation in a 'keyphrase-guided' or perhaps more commonly known as the 'answer-focused' format, where the inputs are a given "context" and a "keyphrase" within that context. Traditionally, this "keyphrase" is assumed to be the answer to the question that is generated. This is as opposed to the end-to-end format, where a body of text is given as input and a set of questions are produced as output, with or without the answers. We pursue this avenue since we believe that this is much more controllable, relevant and much more deployable in a real world application. 

Another contribution of this paper is the improvement on the coherence and relatedness of the answers with this new adversarial approach, since the answer is generated by the question answering system, given the question, instead of being the keyword around which the question was generated. That being said, we find that often the keyword is a subset of the answer span. Due to the fact that many of the generated questions cannot be answered in one phrase, we find that our coherence of answer to question is much higher, since our adversarial system extracts a span of text and is not limited to one phrase. We identify this as a major bottleneck to all question generation systems that have thus far been developed.

In summary, we attempt to solve the following three painpoints in this paper:
\begin{itemize}
    \item We address the problem of scaling, and the problem of possible misinformation generation from question generation systems which is a direct result of the currently most widely used solution to scaling.
    \item We propose an adversarial approach to question generation which includes question answering.
    \item We prove that this approach not only eliminates 'bad' quality questions, but also increases the cohesiveness between the question and the answer
\end{itemize}

\section{Related Works}

In this section, we split the literature review in two parts. We begin the first part by providing a brief history on the development of question-generation models. We then follow it by introducing a concise discussion over the impact of AI-generated question-answering models on the modern society.

\subsection{Understanding Question Generation Models}

While there have been several earlier works on question generation, \cite{heilman2010good, brown2005automatic} most are grammar-focused methods for question generation, where rule-based techniques are employed, while others are question pattern focused, where commonly used question patterns are mined from large text corpora and re-used. With the advent of sequence-to-sequence models, there has been renewed interest in question generation as a field. We have seen several attempts at question generation using RNN-based sequence-to-sequence models, like \cite{wang-etal-2020-pathqg} where the authors have created an end-to-end RNN based question generator, with query representation learning applied to query-based question generation. 

There is also RNN-based question generation based on knowledge graphs \cite{reddy-etal-2017-generating}. Interestingly, \cite{duan-etal-2017-question} gives two types of distinct question generation approaches, one using a CNN and a retrieval based techniques, and another using and RNN and a generation based mechanism. In their work, they attempt rather the opposite of what we attempt in that they attempt to use question generation systems to increase the performance of question answering systems. We however, use a question answering system to further increase the quality of the question generation system. Interestingly, most of the methods we have seen so far are neglecting the scaling problem almost completely and instead chooses to focus on the quality of the generated question. Since we do not find explicit solutions to the scaling problem before the transformer era, we can only hypothesize that it was not imagined to be a problem due to the fact that RNNs can take arbitrarily long sentences. Researchers have noted decreased performance over larger sequences, but this problem is a generalized problem for all RNNs.\newline  With the advent of transformers \cite{vaswani2017attention} that we find several approaches using the encoder-decoder architecture for question generation. It is at this point that the scaling problem comes into the picture, since transformers, unlike RNNs, have a limited input capacity, and therefore text has to be chunkified before being given as input. Since this is a extremely expansive list with applications across domains like visual question generation, and since there are dedicated surveys to cover this topic, we refer the reader to \cite{pan2019recent, kurdi2020systematic,ch2018automatic, das2021automatic} for a very broad list of all the various implementations of transformers with applications to question generation. Within this broad research area, we identify answer-guided transformer-based question generation as being the best performing, most widely usable, and most relevant to the real world. \cite{lopez2020transformer} is a transformer-based end-to-end question generation approach. They use GPT-2 \cite{radford2019language} for the actual generation process. There have also been BERT \cite{devlin2018bert} based models for question generation, given by the work in \cite{chan-fan-2019-recurrent}, where the authors developed a sequential model for BERT based question generation. However, these works and results are compelling cases for using a full-transformer for this task instead of just an encoder or just a decoder stack. Due to the performance demonstrated in \cite{10.1145/3442381.3449892, raffel2020exploring}, and since a full transformer is an encoder-decoder network, instead of just an encoder stack or a decoder stack, we choose to use a T5 \cite{raffel2020exploring} for the generation process. Going by the results given in \cite{10.1145/3442381.3449892}, we realize that a well-trained Pegasus \cite{zhang2020pegasus} could give a comparable performance, but we leave those studies up to future research. Pretrained transformers for question generation tasks give state of the art performance across a wide range of domains. \newline We also see a progression towards question generation with multiple choice answers \cite{10.1145/3442381.3449892}, where the authors also used a T5 and Pegasus models for generation, and then created distractor options for a given answer. This line of work is perhaps the closest to what we are doing now.
However, most critically, they use summarizers to scale a given text to fit the input size of a transformer, and further, they take the keyword as an answer as a given. In our approach, the answer is regenerated using a question answering system. 

We also find that there is considerable research in terms of controllable generative models \cite{hazarika2021zero, keskar2019ctrl}. However, for question generation at the moment, there exists no objective metrics by which we can judge the semantic quality of a question. Although there are "syntactical" metrics, like ROUGE \cite{lin-2004-rouge} and METEOR \cite{banerjee2005meteor}, these scores need references to test how good the produced output is. For question generation, the same question can be asked in a variety of different ways. Question quality prediction remains an open research topic. For these reasons, even though there is research in the direction of controlled language models, we cannot leverage them to further fine-tune question generation.
There is also significant research in terms of question generation in the domain of computer vision \cite{krishna2019information}, some with question answering integrated into it \cite{li2018visual} and other research where it is goal-oriented \cite{zhang2018goal}. However, visual question generation is outside the scope of this paper.
Finally, we find that in most of these approaches, especially in the transformer-based approaches, researchers have used abstractive summarization to shorten the length of the input to make it scale. Critically, fake news generation is a known problem with abstractive summarization models \cite{shu2020combating,esmaeilzadeh2019neural}. While there has been research towards the analysis of factual statements made from abstractive summarizers \cite{lux2020truth, kumar2022textminor}, none of these approaches are ready to be integrated with tasks such as question generation.

\subsection{Understanding the Social Impact of AI-Generated Question-Answering}

AI-Generated content is prone to error, with or without malicious intent from the content creators. In fact, AI-generated bots and content generators have already incurred solid repercussions in the political participation and engagement \cite{kreps2022all}. As even real credible users may spread false information and disinformation without good judgment \cite{shu2019beyond}, there is no guarantee that AI-generated contents, even learned from absolutely human-created resources, are trustworthy \cite{yampolskiy2016taxonomy}, not to mention the malicious acts \cite{zellers2019defending}. 

While there are many powerful methods we can apply to tackle misinformation and disinformation, such as AnswerFact \cite{zhang2020answerfact}, EANN \cite{wang-etal-2020-pathqg}, Mvae \cite{khattar2019mvae}, and many else \cite{shu2019beyond, lu2020gcan}, none of them are quality-driven. Moreover, many recent methodologies takes a significant infrastructure toll on latency, model complexity and computing resources. Note that many public institutions and agencies cannot afford to train a large AI model for their in-house texts, and applying an unknown QA model can be very risky. As discussed in the earlier of this writing, relying on human-in-the-loop detection is post-hoc and dangerous.

We aim to develop an industry-strength, light-weight question generation framework that can be safely scaled to massive amount of content. In the next two sections, we introduce how the new pipeline applies to news aggregation and news ranking applications with high confidence of generated questions.

\section{Problem Definition and Study Procedure}
The objective in question generation is to generate a question from a given body of text. There are two widely accepted approaches to do this. One is to employ "end-to-end" question generation systems \cite{10.1145/3442381.3449892}, where a set of questions are simply generated from a given body of text, with no additional inputs, or control over the generation process. The second approach is "answer-guided" question generation, where the answer is given along with the "context" from which to generate the question. This is the approach we are pursuing in this paper, as it provides significantly more control over the generation process. 
\subsection{Scaling and Misinformation}
Due to the size limitation problem of modern transformers, the most widely used approach to scale a given question generation system is to use a summarization model, which is usually another large language model \cite{zhang2020pegasus} which has been fine-tuned for the summarization task. We find in our experiments that this yields misinformation due to the inherent biases in any large language models that are developed during training. 

\begin{figure}[h]
  \centering
  \includegraphics[width=\linewidth]{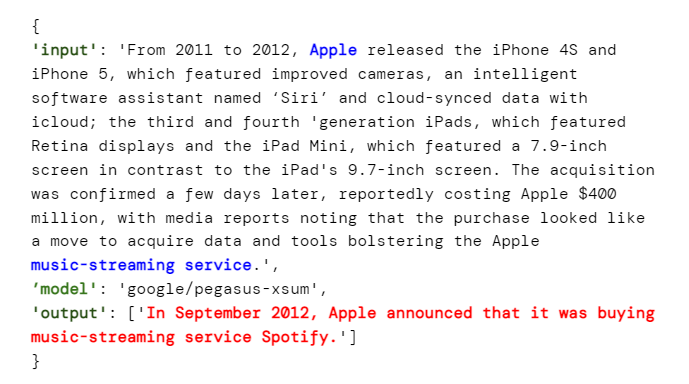}
  \caption{Fake News generated from a State-of-the-Art Summarization model.}
  \label{fig2}
\end{figure}
We define Maximal Generation as the process of generating a question-answer pair for every keyword-context pair. We use this approach to solve the scaling problem, instead of using a summarization model or subsampling the data.
\begin{figure*}[h]
  \centering
  \includegraphics[width=\linewidth]{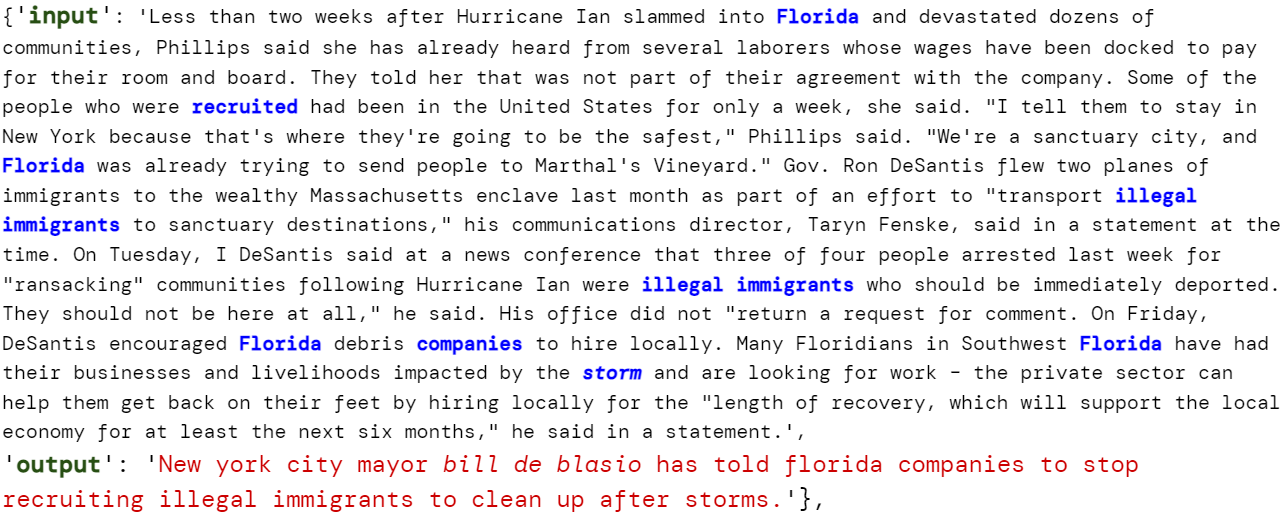}
  \caption{Fake News generated from a Pegasus XSUM. The sheer 'fakeness' is comical.}
\end{figure*}
We employ KeyBERT \cite{sharma2019self} for keyphrase extraction. This does mean that the number of questions that are generated for a given body of text is more than the number of keywords. Given that the question-answer pair is being generated for each keyword-sentence pair, we also find experimentally that in all of the cases we have seen so far, the input dimension of a transformer is large enough to not cause a problem. While it is true that we do not summarize, we also find experimentally that very often, summarization includes significant downsampling in terms of the amount of information that is being conveyed in the output phrase. In practice, this same result can be achieved by simply discarding several of the questions that are generated. Indeed it is to prevent misinformation and semantic corruption that we do not summarize or downsample the amount of information we are presented with.

\subsection{Data}

We collected data from our top push news articles database in which each articles have at least more than 500 words and have more than 1,500 user clicks on push notifications on such articles. The dataset consists of 11,028 news articles, which well represents current trend of user read habits on news articles. To this end, we employed a segmented sampling of different news article categories that intend to include articles from a broad array of questions. For the majority of them (6,053) we have some hashtags associated with the article. We excluded articles that are heavily incentivized or intent for advertisement (mostly in food \& drinks section). A small percentage ({$\sim$ 3.5\%}, 212) of them are considered 'evergreen' content which has a lifespan much longer than typical news articles. The purposes of including such golden set is two fold: (1) we use them for a qualitatively assessed dataset with human-in-the-loop judgments and quality measurement (as discussed in Section 3.4); (2) we used them to verify the integration of generated questions into the news knowledge graph, and the effectiveness of building a 'question graph' in which questions with relevant entities are attached incoming articles.

The top push articles are stored in an enterprise-internal content platform and updated hourly and to be retrieved from Hive. For the purpose of pipeline training, we use the past seven days of data. The knowledge graph is built in Neo4j, with external (Wiki) entities removed to exclude confounding effects from the question answering generation framework.

\subsection{Adversarial Question Generation Pipeline}
\begin{figure*}[h]
  \centering
  \includegraphics[width=\linewidth]{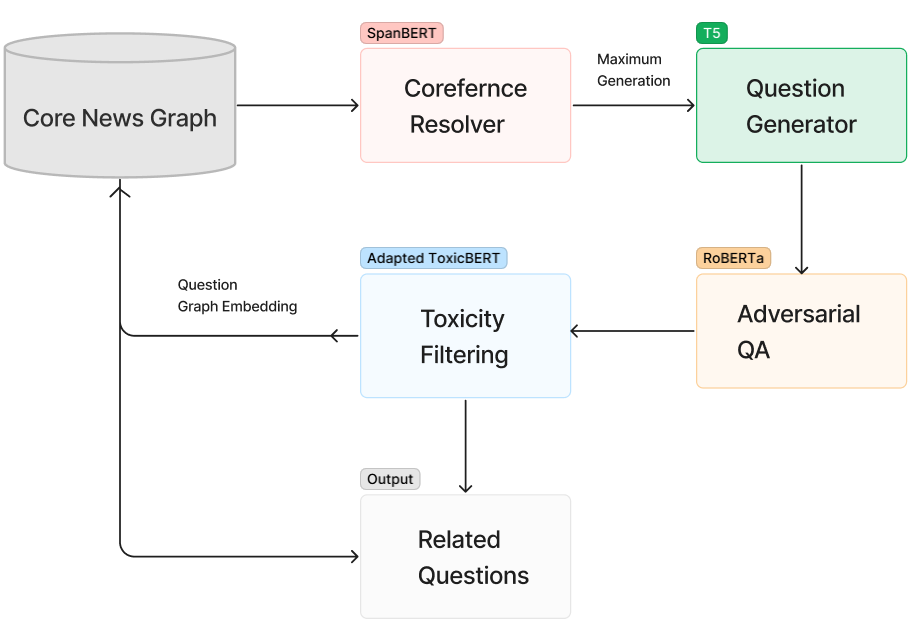}
  \caption{QA Gen Pipeline with News KG}
  \label{fig:pip}
\end{figure*}

Our adversarial question generation pipeline includes four steps. An overview of this pipeline is available in Figure~\ref{fig:pip}.

The first step is where we employ a coreference resolver \cite{joshi-etal-2020-spanbert} to tackle all the coreferences within a a given body of text to allow questions to better stand on their own.
\begin{figure}[h]
  \centering
  \includegraphics[width=\linewidth]{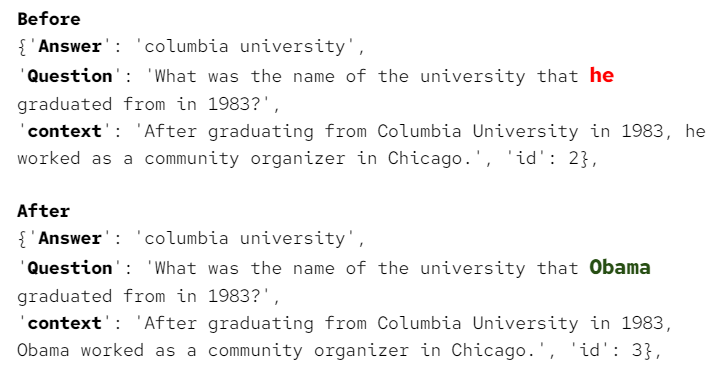}
  \caption{Before and after coreference resolution for question generation}
  \label{fig:cor}

\end{figure}
To our knowledge, we are also the first to attempt to use coreference resolution \cite{ng2002improving} to improve the quality of a question generation system. Experimentally, we notice that the impact of this is extremely dependant on the input text and the inherent limitations on the performance of the coreference resolver. As shown in Figure~\ref{fig:cor}, we substantially corrected the problem. For this reason, we do not conduct an extended study on the use of coreference resolution for question generation. We leave optimizations of this performance up to future research. 

For Step 2, we use a KeyBERT for keyphrase extraction from a given context. For each sentence in which the extracted keyword is relevant, we extract that sentence to obtain a set of context-keyword pairs. Each of these context-keyword pairs, we fine-tuned a T5-large \cite{raffel2020exploring} (NewsGraph-T5) trained on our dataset to generate a question that are specifically fit in the news feed recommendation setting.

In Step 3, we generate an answer, instead of taking the keyword as the default answer like the current approaches, using a fine-tuned RoBERTa model, which has shown to give state-of-the-art results in question answering. Note that questions generated at this step is matched to the news knowledge graph, where other relevant questions are searched and queried for to congregate the best predicted quality of question sets for a given article. Linking question-answering framework will substantially amplify the quality of the question-answering framework performance. However, to exclude confounding effects, all the statistics in the 'Results' section will not include the additional benefits from knowledge graph. In other words, we consider only the questions being generated from a single instance.

As for the last step, since the primary RoBERTa model we are using is trained on SQuaDv2.0 \cite{rajpurkar2018know} which contains negative examples, we filter out any question-context pairs for which the model gives '<s>' output which is the delimiter character. In order to crack down any questions include profanity, slurs, and discriminative wordings, we fine-tuned a ToxicBERT \cite{luu2021uit} model for filtering out any toxicity in the generated output without harming the legitimate questions that include some words with intensity. Since ToxicBERT is an established system in and of itself, we do not conduct further reasoning to profile its performance.

We are attempting to answer a given question using a question answering model which has been trained on superior data that include open questions. The reason we cannot use this same superior data to further constrain the generation process for question generation is that constraining the generation process with unanswerable questions degenerates into a much large and much more theoretical problem in modern artificial intelligence, which is constrained generative models. We find that for question generation this approach is more appropriate since we only get questions that can be answered from a given context, and does not rely on assumed inherent knowledge on the part of the reader. However, this non-support does not disqualify us from achieving our goal, as we are aiming for a pipeline that generates reliable, safely scaled questions that can be traced from hard evidence in the articles. Open-source answering invites too many unknown variables, and thus is outside of the scope for this work.

\subsection{Evaluation and Focused User Studies}
Evaluation of question-answering quality is extremely intricate. In light of standard evaluation metrics for a question-answering metrics that focused on \textit{accurate answering}, metrics like mean reciprocal ranking \cite{wu2011optimizing}, F-1 score or exact matching make sense. However, a common theme of MRR, F-1 score and EMs are that they all reward \textit{brevity}. Brevity on its own is not bad. However, overemphasis on brevity has made question answering framework producing shorter, hard-to-digest information than they could have been to serve readers. Readers typically enjoy complete sentences and a comfortable-sized answer that provide them answers that require least amount of effort. Therefore, we propose two qualitative evaluations of the question-answering qualities of our model compared to the others. 

\textit{User Surveys and Focused Studies} We conduct one user survey and one user focused studies, respectively on perceived question generation quality for the top selected questions and the distribution of qualities for \textit{all} questions being generated from our system. The details regarding how to conduct the user survey and focused studies are included in Section 4.

Readers may well challenge us for why we do not include quantitative measures anyway. We do not feel including EMs, F-1 score or MRR is a fair game because in the following examples, we focus on \textit{generating questions} rather than merely answering preset questions. There is no groundtruth, even for a public dataset to evaluate other models for the questions we generate. We further argue that the other alternatives (XLM-Roberta, tinyBERT, etc.) can be easily plugged into our pipeline. We use the current fine-tuned RoBERTa given our best interests in training resources, but other options are also supported.

\section{Experiments and Results}
We conduct two sets of experiments. The first is where we profile the performance of an adversarial system v/s a vanilla question generation system. The second set of experiments is for our scaling approach, where we attempt to profile the differences between scaling using abstractive summarization vs our "Maximal Generation" scaling approach. The advantages of having an adversarial approach include an increased answer quality for a given question. One hypothesis is that the answering system will filter out unanswerable questions due to the fact that it was trained on a superior dataset, as opposed to vanilla methods, where there is no filtering involved. We also hypothesise that this approach will increase the coherence between a question and its answer. Secondly we analyze the impact our scaling approach has on the question generation pipeline, as opposed to the traditional approaches to scaling which is to use abstractive summarization and/or to downsample the amount of data that is being fed into the generator. We are not concerned about the performance of KeyBERT for keyphrase extraction and therefore we avoid questions created from irrelevant keywords.

\subsection{Adversarial System}
First we test the hypothesis that an adversarial system gives a much lower number of unanswerable questions by filtering out unanswerable questions that are generated. To do this, we generate 500 questions from a vanilla T5 question generator, which has shown to give state-of-the-art performance previously, and then we use the exact same keyword-context pair to generate questions using our adversarial approach. We filter out questions for which the answering systems gives the special token "<s>" output. We record the "Dropout", which is defined as the fraction of questions that are thrown away from the generated output by the QA system.
\begin{figure}[h]
  \centering
  \includegraphics[width=\linewidth]{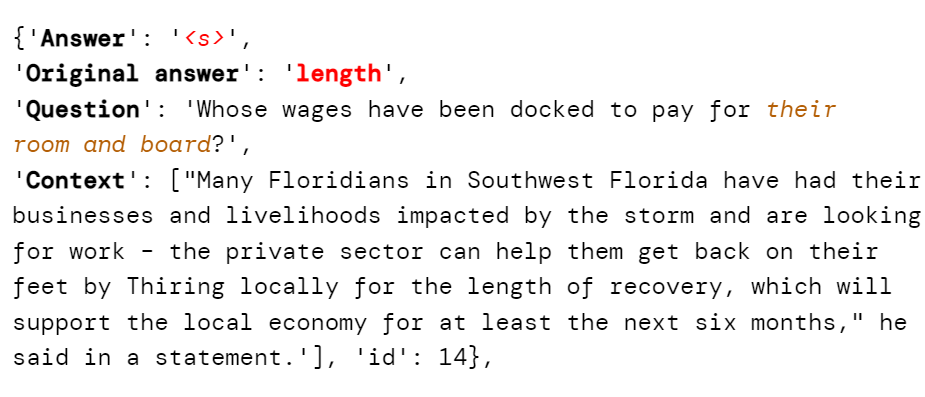}
  \caption{A non-sensical question that has been generated and being flagged with the '<s>' token by the QA module.}
\end{figure}
\begin{figure}[h]
  \centering
  \includegraphics[width=\linewidth]{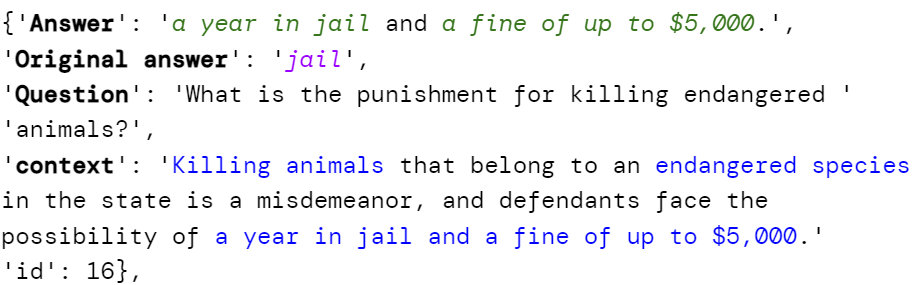}
  \caption{The answer given by our adversarial approach is significantly better than the vanilla answer used in the traditional approach.}
\end{figure}
To further understand these results, we do a human evaluation on the quality of the generated questions, on a 1-5 Likert Scale. The human study is done by a group of 168 graduate students, product managers, data scientists, and software engineers, selected at random, and are asked to rate the quality of a question without the answer; at this stage, we are simply using the QA model to filter out bad questions. Since we are using more than 100 news articles to generate the questions, we do notice that the inherent quality of the article has an impact on the dropout rate, but in this experiment, we are observing that the average quality of the questions independent of factors like the relevance of the input article, keyword, etc is higher with our adversarial approach (Results given in Table 1). This is because poorer quality generated questions are discarded by the answering system.

Secondly, we use the same experimental methodology to measure the "Coherence" between the question and answer, in both the vanilla approach and our adversarial approach. "Coherence" is defined as the explainability of a given answer to a given question; \textit{How well does an answer suit a given question?}. For a question-answer pair to achieve 5 on the Likert scale would mean that the answer is a perfect explanation to the question and that the human needs no further clarification on the matter. This is also given in Table 1. Moving on to the "Quality" metric, to elaborate on the experimental setup regarding the quality of a question, we ask the participants to score the question regardless of the relevance of the topic. For instance, a question that requires the reader to have some context knowledge, given which, the question is understandable and is perfect in every other way like syntax and grammar, is scored very close to 5 on the scale. Each question is scored by at least two participants.
\begin{table}[h]
\centering
\begin{tabular}{|l|l|l|l|}
\hline
 & Coherence & Dropout & Quality \\ \hline
\begin{tabular}[c]{@{}l@{}}Vanilla Question\\ Generator\end{tabular}                       & 3.191  & N/A    & 2.542  \\ \hline
\begin{tabular}[c]{@{}l@{}}Adversarial Approach\\ (RoBERTa-Base-SQuaD2)\end{tabular}       & \textbf{3.810} & 22\%   & \textbf{3.661} \\ \hline
\begin{tabular}[c]{@{}l@{}}Adversarial Approach\\ (RoBERTa-Base-SQuaD1.1)\end{tabular}     & 3.577 & 18.6\% & 3.161 \\ \hline
\begin{tabular}[c]{@{}l@{}}Adversarial Approach\\ (BERT-Large-Uncased-SQuaD2)\end{tabular} & 3.673 & 26.8\% & 3.161 \\ \hline
\end{tabular}
\caption{Experimental Results for Vanilla approach vs Adversarial approach with various models.}
\label{tab:Table 1}
\end{table}

We do not consider the relevance of a given news article in this experiment, since we are looking to objectively judge the performance differences between a vanilla generator, and our adversarial approach. Furthermore, the keyword extractor has an impact on the relevance of the question-answer pair.

We consider three models, two of which are RoBERTa variants, and the last one is a BERT-Large-Uncased. One of the RoBERTa models and the BERT model is trained on SQuaDv2.0 \cite{rajpurkar2018know}, and the remaining RoBERTa variant is trained on SQuaDv1.1. We see, rather surprisingly, that the RoBERTa model trained on SQuaDv1.1 is not far behind the same model trained on SQuaDv2.0, although SQuaDv2.0 has negative examples specifically designed for reading comprehension. Rather predictably, the vanilla BERT model falls behind both the RoBERTa variants, except in the "Dropout" criterion. Interestingly, despite dropping out a larger fraction of the question, we see that the quality does not improve much (or at all). This does go to show the relevance of the RoBERTa model for question answering.
We notice the impact of source quality on dropout since the dropout scores are relatively lower than text taken from relatively more famous sources.

Since we are not employing abstractive summarization or downsampling to scale our system, we control the number of questions that have been generated by limiting the number of keywords that are extracted from every article. In our experiments, since the article quality may vary, and since we are not attempting to constrain the fundamental question generator, we discard non-sensical questions when measuring the coherence of the question-answer pair. Another subject we address is the degenerative nature of over-generalized questions. For example, a question along the lines of "What type of wolves are being listed as endangered by the State?" cannot be answered without its accompanying context. However, since neither the generator nor the answering system can be faulted for this, we do consider this as an acceptable question since, since the objective is to measure the syntactic and semantic coherence and quality of a question within the context.

\subsection{Maximal Generation versus Abstractive Summarization}
We define Maximal Generation as the process of generating a question-answer pair for every keyword-context pair. We conducted an experiment where we used abstractive summarization and generated questions versus when we used our approach to generate questions. We measure three metrics, namely "Articles Needed", which is the number of articles needed to generate 100 questions, "Relevance Fraction", which is the number of questions that are relevant from the generated set, and "Quality"; measured in the same way as in the previous experiment, using the same adversarial approach. For measuring the number of articles needed, we use the "Pegasus-XSum" abstractive summarization model \cite{zhang2020pegasus} to create a summary of the input article before generating questions based on that. Due to size limitations, we split up a news article into chunks as needed, where the maximum length is set to be less than 512, splitting at the closest sentence.
\begin{table}[h]
\centering
\begin{tabular}{|l|l|l|l|}
\hline
                                                                               & \begin{tabular}[c]{@{}l@{}}Articles Needed\\ (Per 100 Questions)\end{tabular} & \begin{tabular}[c]{@{}l@{}}Relevance \\ Fraction\end{tabular} & Quality \\ \hline
\begin{tabular}[c]{@{}l@{}}Abstractive\\ Summarization\end{tabular}            & 42.2                                                                          & 0.82                                                          & 3.425    \\ \hline
\begin{tabular}[c]{@{}l@{}}Maximal \\ Generation\\ (Our approach)\end{tabular} & \textbf{6.67}                                                                          & \textbf{0.15}                                                          & \textbf{3.3}    \\ \hline
\end{tabular}
\caption{Experimental Results for Abstractive Summarization vs Maximal Generation}
\label{tab:Table 2}
\end{table}
We realize that chunkification has an impact on the number of questions per article, and that these results will vary across specific models used, especially if the summarization is done using long-document transformers \cite{beltagy2020longformer}. However, irrespective of the specific model used (state-of-the-art Pegasus model in our case), we see that the number of questions generated by our adversarial approach is significantly more than the number of questions generated by the abstractive summarization approach. While the number of chunks affects the number of questions generated in the vanilla approach, our scaling approach does away with this limitation. This is because usually, we find that abstractive summarization only gives 1-2 sentences as a summary, while we generate a question-answer pair for each keyword-sentence pair. Since our approach does not have this limitation, the number of questions we generate is independent of the number of chunks the text is split into. Moving on to "Relevance Fraction", this is the number of questions that are retained from the generated question set, judging by the relevance of the question topic with respect to the source article topic.

We find that maximal generation did generate a set of irrelevant questions since the keyword extraction model (KeyBERT) identifies some keyphrases as relevant. That said, we see that the relevance fraction for maximal generation is significantly lower since we find that only the top few questions are relevant within the context of the article. Other questions do not tend to be in keeping with the context. However, on several occasions, we did find that abstractive summarization has given fake news. To our knowledge, none of the generated questions from our approach are factually incorrect.

\begin{figure}[h]
  \centering
  \includegraphics[width=8.8cm]{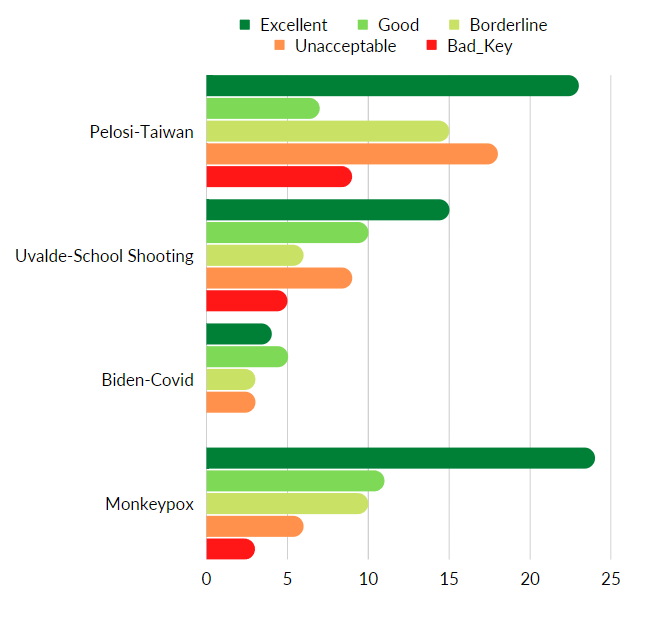}
  \caption{Results of the Perceived Quality of ALL questions generated from the same article}
  \label{fig:per}
\end{figure}

The final measurement is with respect to the quality, which is measured in the same way as it was in the previous experiment. We see that the quality is not affected much by the scaling, as it has very little to do with the actual generation process. We do see a slight increase for the abstractive summarization technique, and we hypothesize that this is due to the fewer number of questions that are extracted from each article, thereby showing an increased quality. Note that despite the larger number of questions being generated, the perceived quality of generated questions are overwhelmingly good. Here, we invite anonymous reviewers to evaluate and label the quality of \textbf{all} questions being generated from the \textit{same} article. To avoid biases, each reviewer is given only one random article with all of its generated questions at a time. We demonstrate the results in Figure~\ref{fig:per}. As we can observe here, the overall perceived quality of questions are still very good: the majority of questions are not oversimplified, incomprehensible or of less reasonable interests.

In short our experimental results gives the following conclusions:
\begin{itemize}
    \item Our scaling works does not generate fake news, since it does not involve creating extra data.
    \item The "Coherence" and "Quality" of generated questions are higher with our adversarial approach, since all un-answerable questions and trivial questions with too many answers are filtered out.
    \item The number of articles we need to generate a given number of questions is much lower with our scaling approach.
    \item While our scaling approach does produce irrelevant questions, we do end up with more relevant questions per 100 questions than vanilla approaches.
\end{itemize}
We also have a ToxicBERT model to prevent any toxic output from the pipeline. However, since it is a previously well-established model \cite{luu2021uit}, we do not conduct further experiments on it.






\section{Future Research and Current Limitations}
We leave larger ablation studies like replacing the Question Answering system that we are currently using with higher performance longformer \cite{beltagy2020longformer} models to future research since in this paper, our objective was to simply show that an adversarial approach gives better generative quality for question generation; although we did consider several of the state-of-the-art models, we do not analyze any QA model of the longformer family. Another objective that we had in this paper is to show that we do not need to have abstractive summarization which can cause misinformation and we do not need to subsample the data. While we showed that our approach to scaling decreases the amount of data required to generate any number of given questions, we were unable to do larger studies on the flat-lining of the quality versus the number of keywords extracted. Current limitations include the lack of studies done on the quality of generated questions, the absence of purely objective measures, and the absence of any \textit{sufficiently large groundtruth data} (except our golden dataset, which is not open) to judge the quality of a question outside of a given context. Given these limitations, question generation and quality analysis for questions becomes a very under studied topic. Therefore for future research, we propose the creation of a dataset, which includes relevance, as calculated from a topic graph. Furthermore, we propose training PLMs with objectives that promote the learning of the topic relevance and bias towards generating questions that can be assumed to have a high relevance and quality. Finally, we also propose the finetuning of PLM encoders to identify the difference between high syntactical and low syntactical question quality.

\bibliographystyle{alpha}
\bibliography{main}

\newcommand{\etalchar}[1]{$^{#1}$}
\begin{thebibliography}{WWF{\etalchar{+}}20}

\bibitem[ANKM22]{agarwal2022deep}
Rajat Agarwal, Vaishnav Negi, Akshat Kalra, and Ankush Mittal.
\newblock Deep learning and linguistic feature based automatic multiple choice
  question generation from text.
\newblock In {\em International Conference on Distributed Computing and
  Internet Technology}, pages 260--264. Springer, 2022.

\bibitem[ASL20]{angeline2020can}
Mia Angeline, Yuanita Safitri, and Amia Luthfia.
\newblock Can the damage be undone? analyzing misinformation during covid-19
  outbreak in indonesia.
\newblock In {\em 2020 International Conference on Information Management and
  Technology (ICIMTech)}, pages 360--364. IEEE, 2020.

\bibitem[BFE05]{brown2005automatic}
Jonathan Brown, Gwen Frishkoff, and Maxine Eskenazi.
\newblock Automatic question generation for vocabulary assessment.
\newblock In {\em Proceedings of Human Language Technology Conference and
  Conference on Empirical Methods in Natural Language Processing}, pages
  819--826, 2005.

\bibitem[BL05]{banerjee2005meteor}
Satanjeev Banerjee and Alon Lavie.
\newblock Meteor: An automatic metric for mt evaluation with improved
  correlation with human judgments.
\newblock In {\em Proceedings of the acl workshop on intrinsic and extrinsic
  evaluation measures for machine translation and/or summarization}, pages
  65--72, 2005.

\bibitem[BPC20]{beltagy2020longformer}
Iz~Beltagy, Matthew~E Peters, and Arman Cohan.
\newblock Longformer: The long-document transformer.
\newblock {\em arXiv preprint arXiv:2004.05150}, 2020.

\bibitem[CF19]{chan-fan-2019-recurrent}
Ying-Hong Chan and Yao-Chung Fan.
\newblock A recurrent {BERT}-based model for question generation.
\newblock In {\em Proceedings of the 2nd Workshop on Machine Reading for
  Question Answering}, pages 154--162, Hong Kong, China, November 2019.
  Association for Computational Linguistics.

\bibitem[CS18]{ch2018automatic}
Dhawaleswar~Rao Ch and Sujan~Kumar Saha.
\newblock Automatic multiple choice question generation from text: A survey.
\newblock {\em IEEE Transactions on Learning Technologies}, 13(1):14--25, 2018.

\bibitem[DCLT18]{devlin2018bert}
Jacob Devlin, Ming-Wei Chang, Kenton Lee, and Kristina Toutanova.
\newblock Bert: Pre-training of deep bidirectional transformers for language
  understanding.
\newblock {\em arXiv preprint arXiv:1810.04805}, 2018.

\bibitem[DMPS21]{das2021automatic}
Bidyut Das, Mukta Majumder, Santanu Phadikar, and Arif~Ahmed Sekh.
\newblock Automatic question generation and answer assessment: a survey.
\newblock {\em Research and Practice in Technology Enhanced Learning},
  16(1):1--15, 2021.

\bibitem[DSC17]{du2017learning}
Xinya Du, Junru Shao, and Claire Cardie.
\newblock Learning to ask: Neural question generation for reading
  comprehension.
\newblock {\em arXiv preprint arXiv:1705.00106}, 2017.

\bibitem[DTCZ17]{duan-etal-2017-question}
Nan Duan, Duyu Tang, Peng Chen, and Ming Zhou.
\newblock Question generation for question answering.
\newblock In {\em Proceedings of the 2017 Conference on Empirical Methods in
  Natural Language Processing}, pages 866--874, Copenhagen, Denmark, September
  2017. Association for Computational Linguistics.

\bibitem[EPX19]{esmaeilzadeh2019neural}
Soheil Esmaeilzadeh, Gao~Xian Peh, and Angela Xu.
\newblock Neural abstractive text summarization and fake news detection.
\newblock {\em arXiv preprint arXiv:1904.00788}, 2019.

\bibitem[HNHT21]{hazarika2021zero}
Devamanyu Hazarika, Mahdi Namazifar, and Dilek Hakkani-T{\"u}r.
\newblock Zero-shot controlled generation with encoder-decoder transformers.
\newblock {\em arXiv preprint arXiv:2106.06411}, 2021.

\bibitem[HS10]{heilman2010good}
Michael Heilman and Noah~A Smith.
\newblock Good question! statistical ranking for question generation.
\newblock In {\em Human Language Technologies: The 2010 Annual Conference of
  the North American Chapter of the Association for Computational Linguistics},
  pages 609--617, 2010.

\bibitem[JCL{\etalchar{+}}20]{joshi-etal-2020-spanbert}
Mandar Joshi, Danqi Chen, Yinhan Liu, Daniel~S. Weld, Luke Zettlemoyer, and
  Omer Levy.
\newblock {S}pan{BERT}: Improving pre-training by representing and predicting
  spans.
\newblock {\em Transactions of the Association for Computational Linguistics},
  8:64--77, 2020.

\bibitem[JZ20]{jerit2020political}
Jennifer Jerit and Yangzi Zhao.
\newblock Political misinformation.
\newblock {\em Annual Review of Political Science}, 23(1):77--94, 2020.

\bibitem[KBFF19]{krishna2019information}
Ranjay Krishna, Michael Bernstein, and Li~Fei-Fei.
\newblock Information maximizing visual question generation.
\newblock In {\em Proceedings of the IEEE/CVF Conference on Computer Vision and
  Pattern Recognition}, pages 2008--2018, 2019.

\bibitem[KGGV19]{khattar2019mvae}
Dhruv Khattar, Jaipal~Singh Goud, Manish Gupta, and Vasudeva Varma.
\newblock Mvae: Multimodal variational autoencoder for fake news detection.
\newblock In {\em The world wide web conference}, pages 2915--2921, 2019.

\bibitem[KKS22]{kumar2022textminor}
Sujit Kumar, Gaurav Kumar, and Sanasam~Ranbir Singh.
\newblock Textminor at checkthat! 2022: fake news article detection using
  robert.
\newblock {\em Working Notes of CLEF}, 2022.

\bibitem[KLP{\etalchar{+}}20]{kurdi2020systematic}
Ghader Kurdi, Jared Leo, Bijan Parsia, Uli Sattler, and Salam Al-Emari.
\newblock A systematic review of automatic question generation for educational
  purposes.
\newblock {\em International Journal of Artificial Intelligence in Education},
  30(1):121--204, 2020.

\bibitem[KMB22]{kreps2022all}
Sarah Kreps, R~Miles McCain, and Miles Brundage.
\newblock All the news that’s fit to fabricate: Ai-generated text as a tool
  of media misinformation.
\newblock {\em Journal of Experimental Political Science}, 9(1):104--117, 2022.

\bibitem[KMV{\etalchar{+}}19]{keskar2019ctrl}
Nitish~Shirish Keskar, Bryan McCann, Lav~R Varshney, Caiming Xiong, and Richard
  Socher.
\newblock Ctrl: A conditional transformer language model for controllable
  generation.
\newblock {\em arXiv preprint arXiv:1909.05858}, 2019.

\bibitem[LCC{\etalchar{+}}18]{lee2018automatic}
Che-Hao Lee, Tzu-Yu Chen, Liang-Pu Chen, Ping-Che Yang, and Richard Tzong-Han
  Tsai.
\newblock Automatic question generation from children’s stories for companion
  chatbot.
\newblock In {\em 2018 IEEE International Conference on Information Reuse and
  Integration (IRI)}, pages 491--494. IEEE, 2018.

\bibitem[LCCC20]{lopez2020transformer}
Luis~Enrico Lopez, Diane~Kathryn Cruz, Jan Christian~Blaise Cruz, and Charibeth
  Cheng.
\newblock Transformer-based end-to-end question generation.
\newblock {\em arXiv preprint arXiv:2005.01107}, 4, 2020.

\bibitem[LDZ{\etalchar{+}}18]{li2018visual}
Yikang Li, Nan Duan, Bolei Zhou, Xiao Chu, Wanli Ouyang, Xiaogang Wang, and
  Ming Zhou.
\newblock Visual question generation as dual task of visual question answering.
\newblock In {\em Proceedings of the IEEE conference on computer vision and
  pattern recognition}, pages 6116--6124, 2018.

\bibitem[Lin04]{lin-2004-rouge}
Chin-Yew Lin.
\newblock {ROUGE}: A package for automatic evaluation of summaries.
\newblock In {\em Text Summarization Branches Out}, pages 74--81, Barcelona,
  Spain, July 2004. Association for Computational Linguistics.

\bibitem[LL20]{lu2020gcan}
Yi-Ju Lu and Cheng-Te Li.
\newblock Gcan: Graph-aware co-attention networks for explainable fake news
  detection on social media.
\newblock {\em arXiv preprint arXiv:2004.11648}, 2020.

\bibitem[LN21]{luu2021uit}
Son~T Luu and Ngan Luu-Thuy Nguyen.
\newblock Uit-ise-nlp at semeval-2021 task 5: Toxic spans detection with
  bilstm-crf and toxicbert comment classification.
\newblock {\em arXiv preprint arXiv:2104.10100}, 2021.

\bibitem[LSL20]{lux2020truth}
Klaus-Michael Lux, Maya Sappelli, and Martha Larson.
\newblock Truth or error? towards systematic analysis of factual errors in
  abstractive summaries.
\newblock In {\em Proceedings of the First Workshop on Evaluation and
  Comparison of NLP Systems}, pages 1--10, 2020.

\bibitem[LTY21]{10.1145/3442381.3449892}
Adam~D. Lelkes, Vinh~Q. Tran, and Cong Yu.
\newblock Quiz-style question generation for news stories.
\newblock In {\em Proceedings of the Web Conference 2021}, WWW '21, page
  2501–2511, New York, NY, USA, 2021. Association for Computing Machinery.

\bibitem[NC02]{ng2002improving}
Vincent Ng and Claire Cardie.
\newblock Improving machine learning approaches to coreference resolution.
\newblock In {\em Proceedings of the 40th annual meeting of the Association for
  Computational Linguistics}, pages 104--111, 2002.

\bibitem[NO21]{nwafor2021automated}
Chidinma~A Nwafor and Ikechukwu~E Onyenwe.
\newblock An automated multiple-choice question generation using natural
  language processing techniques.
\newblock {\em arXiv preprint arXiv:2103.14757}, 2021.

\bibitem[PLCK19]{pan2019recent}
Liangming Pan, Wenqiang Lei, Tat-Seng Chua, and Min-Yen Kan.
\newblock Recent advances in neural question generation.
\newblock {\em arXiv preprint arXiv:1905.08949}, 2019.

\bibitem[RJL18]{rajpurkar2018know}
Pranav Rajpurkar, Robin Jia, and Percy Liang.
\newblock Know what you don't know: Unanswerable questions for squad.
\newblock {\em arXiv preprint arXiv:1806.03822}, 2018.

\bibitem[RRKJ17]{reddy-etal-2017-generating}
Sathish Reddy, Dinesh Raghu, Mitesh~M. Khapra, and Sachindra Joshi.
\newblock Generating natural language question-answer pairs from a knowledge
  graph using a {RNN} based question generation model.
\newblock In {\em Proceedings of the 15th Conference of the {E}uropean Chapter
  of the Association for Computational Linguistics: Volume 1, Long Papers},
  pages 376--385, Valencia, Spain, April 2017. Association for Computational
  Linguistics.

\bibitem[RSP{\etalchar{+}}21]{roitero2021can}
Kevin Roitero, Michael Soprano, Beatrice Portelli, Massimiliano De~Luise,
  Damiano Spina, Vincenzo~Della Mea, Giuseppe Serra, Stefano Mizzaro, and
  Gianluca Demartini.
\newblock Can the crowd judge truthfulness? a longitudinal study on recent
  misinformation about covid-19.
\newblock {\em Personal and Ubiquitous Computing}, pages 1--31, 2021.

\bibitem[RSR{\etalchar{+}}20]{raffel2020exploring}
Colin Raffel, Noam Shazeer, Adam Roberts, Katherine Lee, Sharan Narang, Michael
  Matena, Yanqi Zhou, Wei Li, Peter~J Liu, et~al.
\newblock Exploring the limits of transfer learning with a unified text-to-text
  transformer.
\newblock {\em J. Mach. Learn. Res.}, 21(140):1--67, 2020.

\bibitem[RWC{\etalchar{+}}19]{radford2019language}
Alec Radford, Jeffrey Wu, Rewon Child, David Luan, Dario Amodei, Ilya
  Sutskever, et~al.
\newblock Language models are unsupervised multitask learners.
\newblock {\em OpenAI blog}, 1(8):9, 2019.

\bibitem[SBA{\etalchar{+}}20]{shu2020combating}
Kai Shu, Amrita Bhattacharjee, Faisal Alatawi, Tahora~H Nazer, Kaize Ding,
  Mansooreh Karami, and Huan Liu.
\newblock Combating disinformation in a social media age.
\newblock {\em Wiley Interdisciplinary Reviews: Data Mining and Knowledge
  Discovery}, 10(6):e1385, 2020.

\bibitem[SL19]{sharma2019self}
Prafull Sharma and Yingbo Li.
\newblock Self-supervised contextual keyword and keyphrase retrieval with
  self-labelling.
\newblock 2019.

\bibitem[SRLB{\etalchar{+}}21]{soprano2021many}
Michael Soprano, Kevin Roitero, David La~Barbera, Davide Ceolin, Damiano Spina,
  Stefano Mizzaro, and Gianluca Demartini.
\newblock The many dimensions of truthfulness: Crowdsourcing misinformation
  assessments on a multidimensional scale.
\newblock {\em Information Processing \& Management}, 58(6):102710, 2021.

\bibitem[STL19]{swire2019public}
Briony Swire-Thompson and David Lazer.
\newblock Public health and online misinformation: challenges and
  recommendations.
\newblock {\em Annual review of public health}, 41:433--451, 2019.

\bibitem[SWL19]{shu2019beyond}
Kai Shu, Suhang Wang, and Huan Liu.
\newblock Beyond news contents: The role of social context for fake news
  detection.
\newblock In {\em Proceedings of the twelfth ACM international conference on
  web search and data mining}, pages 312--320, 2019.

\bibitem[VSP{\etalchar{+}}17]{vaswani2017attention}
Ashish Vaswani, Noam Shazeer, Niki Parmar, Jakob Uszkoreit, Llion Jones,
  Aidan~N Gomez, {\L}ukasz Kaiser, and Illia Polosukhin.
\newblock Attention is all you need.
\newblock {\em Advances in neural information processing systems}, 30, 2017.

\bibitem[WAR18]{wang2018era}
Patrick Wang, Rafael Angarita, and Ilaria Renna.
\newblock Is this the era of misinformation yet: combining social bots and fake
  news to deceive the masses.
\newblock In {\em Companion Proceedings of the The Web Conference 2018}, pages
  1557--1561, 2018.

\bibitem[WB21]{west2021misinformation}
Jevin~D West and Carl~T Bergstrom.
\newblock Misinformation in and about science.
\newblock {\em Proceedings of the National Academy of Sciences},
  118(15):e1912444117, 2021.

\bibitem[WMF{\etalchar{+}}11]{wu2011optimizing}
Yang Wu, Masayuki Mukunoki, Takuya Funatomi, Michihiko Minoh, and Shihong Lao.
\newblock Optimizing mean reciprocal rank for person re-identification.
\newblock In {\em 2011 8th IEEE International Conference on Advanced Video and
  Signal Based Surveillance (AVSS)}, pages 408--413. IEEE, 2011.

\bibitem[WWF{\etalchar{+}}20]{wang-etal-2020-pathqg}
Siyuan Wang, Zhongyu Wei, Zhihao Fan, Zengfeng Huang, Weijian Sun, Qi~Zhang,
  and Xuanjing Huang.
\newblock {P}ath{QG}: Neural question generation from facts.
\newblock In {\em Proceedings of the 2020 Conference on Empirical Methods in
  Natural Language Processing (EMNLP)}, pages 9066--9075, Online, November
  2020. Association for Computational Linguistics.

\bibitem[Yam16]{yampolskiy2016taxonomy}
Roman~V Yampolskiy.
\newblock Taxonomy of pathways to dangerous artificial intelligence.
\newblock In {\em Workshops at the thirtieth AAAI conference on artificial
  intelligence}, 2016.

\bibitem[ZDML20]{zhang2020answerfact}
Wenxuan Zhang, Yang Deng, Jing Ma, and Wai Lam.
\newblock Answerfact: Fact checking in product question answering.
\newblock In {\em Proceedings of the 2020 Conference on Empirical Methods in
  Natural Language Processing (EMNLP)}, pages 2407--2417, 2020.

\bibitem[ZHR{\etalchar{+}}19]{zellers2019defending}
Rowan Zellers, Ari Holtzman, Hannah Rashkin, Yonatan Bisk, Ali Farhadi,
  Franziska Roesner, and Yejin Choi.
\newblock Defending against neural fake news.
\newblock {\em Advances in neural information processing systems}, 32, 2019.

\bibitem[ZHW{\etalchar{+}}22]{zhao2022educational}
Zhenjie Zhao, Yufang Hou, Dakuo Wang, Mo~Yu, Chengzhong Liu, and Xiaojuan Ma.
\newblock Educational question generation of children storybooks via question
  type distribution learning and event-centric summarization.
\newblock {\em arXiv preprint arXiv:2203.14187}, 2022.

\bibitem[ZWS{\etalchar{+}}18]{zhang2018goal}
Junjie Zhang, Qi~Wu, Chunhua Shen, Jian Zhang, Jianfeng Lu, and Anton Van
  Den~Hengel.
\newblock Goal-oriented visual question generation via intermediate rewards.
\newblock In {\em Proceedings of the European Conference on Computer Vision
  (ECCV)}, pages 186--201, 2018.

\bibitem[ZZSL20]{zhang2020pegasus}
Jingqing Zhang, Yao Zhao, Mohammad Saleh, and Peter~J. Liu.
\newblock Pegasus: Pre-training with extracted gap-sentences for abstractive
  summarization, 2020.

\end{thebibliography}

\end{document}